\documentclass[journal]{IEEEtran}

\usepackage{cite}

\ifCLASSINFOpdf
 
\else
 
\fi

\usepackage{amsmath}

\usepackage{graphicx}
\usepackage{resizegather}
\usepackage{tabularx}
\usepackage{caption}
\usepackage{gensymb}
\usepackage{stfloats}
\begin{document}

\title{Scaling Laws for Unamplified Coherent Transmission in Next-generation Short-Reach and Access Networks}

\author{Giuseppe~Rizzelli, Antonino~Nespola, Stefano~Straullu and~Roberto~Gaudino,~\IEEEmembership{Senior Member,~IEEE}% <-this % stops a space
\thanks{G. Rizzelli and R. Gaudino are with the Department
of Electronics and Telecommunications, Politecnico di Torino, Torino,
Italy, e-mail: giuseppe.rizzelli@polito.it.}% <-this % stops a space
\thanks{A. Nespola and S. Straullu are with LINKS Foundation, Torino, Italy.}}% <-this % stops a space
%\thanks{Manuscript received April 19, 2005; revised August 26, 2015.}}

% The paper headers
%\markboth{Journal of \LaTeX\ Class Files,~Vol.~14, No.~8, August~2015}%
%{Shell \MakeLowercase{\textit{et al.}}: Bare Demo of IEEEtran.cls for IEEE Journals}

% make the title area
\maketitle

\begin{abstract}
International standardization bodies (IEEE and ITU-T) working on the evolution of transmission technologies  are still considering traditional direct detection solutions for the most relevant short reach optical link applications, that are Passive Optical Networks (PON) and intra-data center interconnects. Anyway, future jumps towards even higher bit rates per wavelength will require a complete paradigm shift, moving towards coherent technologies. In this paper, we thus study both analytically and experimentally the scaling laws of unamplified coherent transmission in the short-reach communications ecosystems. We believe that, given the extremely tight techno-economic constraints, such a revolutionary transition towards coherent in short-reach first requires a very detailed study of its intrinsic capabilities in largely extending the limitation currently imposed by direct detection systems. To this end, this paper focuses on the ultimate physical layer limitations of unamplified coherent systems in terms of bit rate and power budget. The main parameters of our performance estimation model are extracted through fitting with a set of experimental characterizations and later used as the starting point of a scaling laws study regarding local oscillator power, modulator-induced attenuation, bit rate, and maximum achievable power budget. The analytically predicted performance is then verified through transmission experiments, including a demonstration on a 37-km installed metropolitan dark fiber in the city of Turin. 
Our findings show that coherent detection without optical pre-amplification and using PM-QPSK can tolerate optical power budget (OPB) well above, for instance, the 29 $dB$ imposed by the current PON standards even at extremely high raw bit rates up to 800 $Gbps$. PM-16QAM, on the other hand, can provide up to 190 $Gbps$ at 29 $dB$ OPB only if combined with soft FEC algorithms.  Even higher bit rate are also shown for the less demanding power budget needed in intra-data centers links.
\end{abstract}

% Note that keywords are not normally used for peer review papers.

\IEEEpeerreviewmaketitle

%%%%%%%%%%%%%%%%%%%%%%%%%%%%%%%%%%%%%%%%%%%%%%%%%%%%%%%%%%%%%%%%%%%%%%%%%%%%%%%%%%%%%%%%%%%%%%%%%%%%%%%%%
%%%%%%%%%%%%%%%%%%%%%%%%%%  body  %%%%%%%%%%%%%%%%%%%%%%%%%%

%%%%%%%%%%%%%%%%%%%%%%%%%%%%%%%%%%%%%%%%%%%%%%%%%%%%%%%%%%%%%%%%%%%%%%%%%%%%%%%%%%%%%%%%%%%%%%%%%%%%%%%%%
\begin{figure*}[t!]
\centering
\captionsetup{justification=centering}
\includegraphics[width=0.8\textwidth]{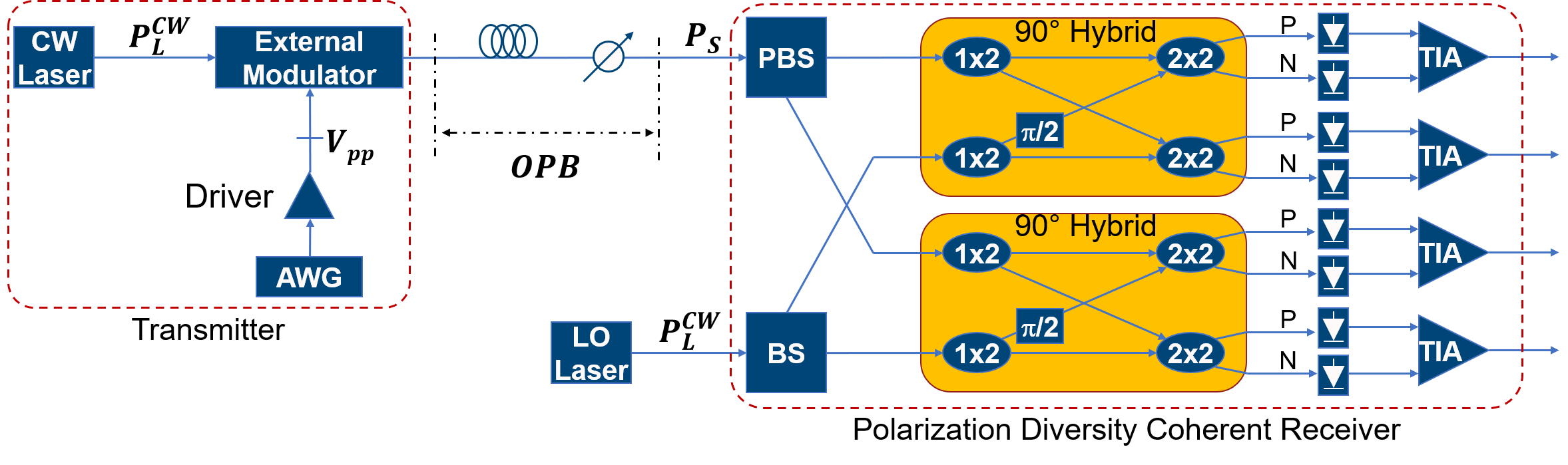}
\caption{Experimental setup of the coherent system. CW: continuous wave. AWG: arbitrary waveform generator. $P_L^{CW}$: continuous wave optical power of the lasers. $V_{pp}$: peak-to-peak voltage. OPB: optical power budget. LO: local oscillator. BS: beam splitter. PBS: polarization beam splitter. $P_s$: received signal optical power. TIA: transimpedance amplifier.}
\label{fig:setup}
\end{figure*}

%%%%%%%%%%%%%%%%%%%%%%%%%%%%%%%%%%%%%%%%%%%%%%%%%%%%%%%%%%%%%%%%%%%%%%%%%%%%%%%%%%%%%%%%%%%%%%%%%%%%%%%%%
\section{Introduction}

\IEEEPARstart{C}{oherent} technology is today the state-of-the-art for any new commercial installation in long-haul fiber systems \cite{Infinera,Winzer}. For techno-economic reasons, coherent is not yet used in the other shorter telecommunication segments, but the situation may rapidly change. For instance, the new standard 400ZR for inter-Data Center Interconnect (DCI) \cite{400G-ZR} is currently introducing coherent also in the metro segment (single span, typically up to 80 $km$), and it is expected to largely reduce the costs of the required transceivers. 

Moreover, also other short distance market segments are today rapidly approaching the typical bit rate limitations imposed by traditional intensity modulation-direct detection (IM-DD) solutions \cite{Kahn1,mecozzi,straullu}, in particular in optical access networks based on PON architectures or intra-DCI links. In this paper, we thus study the ultimate scaling laws of coherent transmission for a potential future use in these short-reach communications scenario, focusing achievable bit rates vs. physical layer parameters.

For what concerns PON, the standardization committees in IEEE and ITU-T are today working towards realizing a new generation of standards for PON transmission at 25 and 50 Gbps per wavelength \cite{IEEE,ITU}, stretching DD systems to their limits, also due to the fact that PON has a very high requirement in terms of optical power budget (OPB) (at least 29 $dB$) \cite{NGPON2}, and of resilience to chromatic dispersion. The next leap in bit rate (100 Gbps per wavelength or more) will thus be increasingly difficult if sticking with DD. 

Similarly, for what concerns DCI, IEEE is rapidly moving towards 100G per wavelength using DD PAM-4, but a jump to even higher bit rate per wavelength relying on DD seems difficult, in this case mostly for the resulting bandwidth limitation in the opto-electronics.

A jump towards coherent technologies may be the next move in both sectors. However, the cost constraints for PON and DCI (particularly for intra-DCI transceivers) are so tight that such a revolutionary paradigm shift may be envisioned only if it ensures a long term and ample roadmap towards ultra high bit rate. We believe that our present study on the physical layer ultimate limitations of coherent in short-reach can be of interest for system vendors and telecom operators, when and if they are willing to consider the new technology.

In this paper, we focus in particular on finding, through analytic formulas and by practical experiments, the maximum bit rates achievable using coherent technologies but without any optical amplification, that is, maintaining the "passive" paradigm of the two main short reach application scenarios (PON and DCI). It is interesting to note that current efforts towards 50G-PON \cite{IEEE} are considering DD systems but only after introducing optical amplifiers (either as boosters or pre-amplifiers, using semiconductor optical amplifiers), while coherent may allow again avoiding optical amplifiers, as shown in this paper, thus partially reducing the cost-gap between DD and coherent transceivers.

We consider here a full implementation of coherent systems having the same transceiver architectures that are today used for long-haul, including all the required Digital Signal Processing (DSP) for phase recovery and dispersion compensation \cite{Pilori}, while we do not consider "ad-hoc" simplified solutions for coherent over short-reach which are largely present in the literature \cite{Erkilinc,Matsuda}, but  are often not suitable for  target bit rates well above 100 Gbps per wavelength and/or for high OPB. However, for intra-data center applications, with target length up to 2 km, the accumulated chromatic dispersion can be sufficiently low \cite{Savory} to avoid having a dedicated DSP stage for dispersion compensation, leaving this task to the adaptive equalizer, and thus reducing the ASIC footprint and power consumption. An analysis on cost, complexity and power consumption of the optoelectronic+ASIC is mentioned in the paper conclusions, but it is outside the main scope of this paper. It is anyway worth mentioning that pre-commercial proposal for coherent in short reach starts to appear, as reported in \cite{XR_optics}.

We will show in our study that the unamplified coherent transmission can work well above 100 Gbps per wavelength even at the very high required PON OPB, and even more for DCI. The paper focuses on finding scaling laws towards higher bit rates related to fundamental noise (transmitter Relative Intensity Noise (RIN), thermal noise, shot noise), passive components attenuation and target Bit Error Rate (BER). To this end, we did not consider in the analytical derivation any effect related to opto-electronic bandwidth limitations since we want to focus mostly on physical fundamental limitations (i.e. on the key noise sources), observing that in the history of optical transceivers, any major jump in bit rate (10G, 40G, 100G) has always been achieved by a breakthrough in devices bandwidth. 

Moreover, for what concerns possible application in the PON area, we did not consider yet in our present work any issue related to PON upstream burst-mode transmission \cite{FEC,Drift}. Burst-mode coherent receivers are still an open topic, even though several preliminary works are available in the literature \cite{burst_mode_coherent}. The results presented in this paper thus apply directly to PON downstream transmission, and more in general to PON architectures based on a Point-to-Point (P2P) approach, in which a ultra-high speed Optical Network Unit (ONU), for instance for mobile front-hauling applications, operates on a dedicated pair of wavelengths per each of the two directions, as it was already envisioned for the NG-PON2 standard in its P2P flavour.

We discuss in our paper on the scalability of non-amplified coherent receivers in terms of baud rate, in a short-reach environment, showing novel results for advanced modulation formats, such as PM-16QAM and PM-64QAM, and detailed experimental validation of the analytic formula our work is based upon.

This paper is organized as follows. In Section \ref{sec:Exp} we introduce an analytical performance model of a standard coherent receiver without optical amplification and, in parallel, its experimental confirmation on a setup running PM-QPSK and PM-16QAM transmission at 28 $GBaud$. 
In Subsection \ref{subsec_voltage_opt}, we present the results of the modulation peak-to-peak voltage experimental optimization. By fitting the experimental results with the analytical ones, we extract a set of optimal realistic parameters for the model. In Section \ref{sec:sec_analytical_results} we use these parameters to run a detailed analysis of an amplifier-less coherent short-reach system, showing the scaling laws that describe the maximum achievable OPB as a function of the raw bit rate, for different transmission scenarios. In Section \ref{sec_experimental_validation} we make some considerations on the validity of the analytical results and show the outcome of a practical implementation of the system with transmission experiments on 37-km installed dark fiber. Lastly, in Section \ref{implementation} we make further considerations on the practical implementation of such system and show how varying the main system parameters affects the performance in terms of the maximum admissible OPB.

%%%%%%%%%%%%%%%%%%%%%%%%%%%%%%%%%%%%%%%%%%%%%%%%%%%%%%%%%%%%%%%%%%%%%%%%%%%%%%%%%%%%%%%%%%%%%%%%%%%%%%%%%
\section{Experimental Setup and Analytical Model}
\label{sec:Exp}

Our analysis of coherent for short-reach starts from a set of laboratory experiments that are then used to validate an analytical performance model for unamplified coherent transmission. Fig. \ref{fig:setup} shows our experimental setup comprising an optical modulator to generate PM-QPSK or PM-16QAM signals, a standard single mode fiber (SMF) and a commercial, dual-polarization coherent receiver. We study performance in terms of BER and maximum achievable OPB as a function of several system parameters, such as CW laser power, component attenuation, baud rate and fundamental opto-electronics noise sources but without optical amplification, as typically envisioned for short reach applications. 

All DSP functionalities at the transmitter and receiver side are implemented using the well-known off-line processing approach, including typically used routines such as a Viterbi\&Viterbi algorithm for carrier phase recovery \cite{Savory} and an adaptive equalizer with 2x2 butterfly-structured FIR filters \cite{kikuchi2} for polarization impairments compensation.
In our setup, an external cavity laser (ECL) at the transmitter side generates a continuous wave signal at 1550 $nm$ which is then sent to a Lithium-Niobate dual arm and dual polarization Mach-Zehnder external optical modulator (EOM) operating in single-drive push-pull mode driven by a 92 $GS/s$ arbitrary waveform generator (AWG) working as a four output digital-to-analog converter (DAC) to generate PM-QPSK or PM-16QAM signals. The input digital stream is a pseudorandom binary sequence (PRBS) of degree 15.

The OPB is varied through a variable optical attenuator (VOA), either in a "back-to-back" (B2B) transmission configuration, or over 37 $km$ of metro-network dark fiber installed underneath the city of Turin,  which was intentionally selected in our experiments to check our results in a realistic environment. The  37 $km$ link includes in fact many (relatively old) connectors and crosses several central offices.
The coherent receiver is a commercial one (FIM24706 by Fujitsu) with a 22 $GHz$ 3dB bandwidth. The four electrical signals at the transimpedence amplifiers (TIAs) output are then digitized through a real time oscilloscope serving as an analog-to-digital converter (ADC) running at 200 $GS/s$ and post-processed after proper downsampling at two samples per symbol through an off-line DSP routine in Matlab\textsuperscript{\tiny\textregistered}.

The experiments, as shown in the following, were instrumental to validate the model performance for unamplified coherent receiver reported in \cite{zhang} and confirmed in \cite{kikuchi,lavery}, which we used (with some upgrades) in the following Section \ref{sec:sec_analytical_results} to derive some general scalability laws. We briefly report here the main results of the performance estimation introduced in \cite{zhang}, which derives the equivalent signal-to-noise ratio (SNR) on each of the four output electrical signals (corresponding to each of the four available quadratures), in terms of the relevant system parameters:

\newcommand*{\Sample}{%
SNR_{RX} = \dfrac{P_s}{\dfrac{\sigma^2_{th}}{P_L^{CW}} + P_L^{CW} \cdot \sigma^2_{n_{LO}} \cdot CMRR + \sigma^2_{shot} + \dfrac{P_s}{SNR}_{Q}} \label{eq_SNR}
}
\refstepcounter{equation}
\[
  \resizebox{1\linewidth}{!}{$\displaystyle\Sample$\quad(\theequation)}
\]
where $P_s$ is the received average signal optical power, $P_L^{CW}$ is the continuous wave (CW) optical power of the local oscillator (LO), $CMRR$ is the Common Mode Rejection Ratio of the balanced photodetector, $\sigma^2_{n_{LO}}$ is the variance of the LO RIN contribution, $\sigma^2_{th}$ is the variance of the TIA thermal noise and $\sigma^2_{shot}$ is the variance of the shot noise generated in the photodetection process. These terms can be expressed as:
%
%\begin{align}
%    \sigma^2_{th} = \dfrac{i_{TIA}^2 \cdot B_{eq}^{RX}}{8 \cdot R^2}%
%\end{align}
%\begin{align}
%   \sigma^2_{shot} = \dfrac{q \cdot B_{eq}^{RX}}{2 \cdot R}%
%\end{align}
%\begin{align}
%    \sigma^2_{n_{LO}}  = RIN \cdot \dfrac{B_{eq}^{RX}}{2}%
%\end{align}

\begin{equation}
\sigma^2_{th} = \dfrac{i_{TIA}^2 \cdot B_{eq}^{RX}}{8 \cdot R^2}
%\quad
\text{;}\quad
\sigma^2_{shot} = \dfrac{q \cdot B_{eq}^{RX}}{2 \cdot R}
%\quad
\text{;}\quad
\sigma^2_{n_{LO}}  = RIN \cdot \dfrac{B_{eq}^{RX}}{2}
\end{equation}

where $R$ is the overall responsivity (in [$A/W$]) of the coherent receiver (which includes the coherent receiver passive losses before the photodiodes), $i_{TIA}$ is the input-referred noise current density (IRND in $pA/\sqrt{Hz}$) of a single transimpedance amplifier, $B_{eq}^{RX}$ is the effective noise bandwidth of the receiver, $q$ is the electron charge and $RIN$ is the LO RIN parameter. Finally, the $SNR_Q$ parameter in Eq.(\ref{eq_SNR}) accounts for the additional implementation penalties occurring in a high-speed coherent system such as quantization noise, phase noise, imperfect constellation generation, which are usually signal power-independent effects. 
We added the $SNR_Q$ contribution (which was not present in the model developed in \cite{zhang}) since in most experimental cases we found that the BER curves as a function of the received optical power may exhibit an error floor at low BER values.

Using Eq. (\ref{eq_SNR}) we then derived BER values according to the following well-known Equations for three modulation formats of interest:
\begin{align}
    BER_{QPSK} = \dfrac{1}{2}  \mbox{erfc}\left(\sqrt{\dfrac{SNR_{RX}}{2}}\right)%
    \label{BER_QPSK}
\end{align}
\begin{align}
    BER_{16QAM} = \dfrac{3}{8} \mbox{erfc}\left(\sqrt{\dfrac{SNR_{RX}}{10}}\right)%
    \label{BER_16QAM}
\end{align}
\begin{align}
    BER_{64QAM} = \dfrac{7}{24} \mbox{erfc}\left(\sqrt{\dfrac{SNR_{RX}}{42}}\right)%
    \label{BER_64QAM}
\end{align}

To validate this analytical model for BER performance estimation, we run a set of experimental measurements on the coherent setup of Fig.~\ref{fig:setup} for both PM-QPSK and PM-16QAM modulation and different LO power levels, and then optimized the model parameters to obtain the best fitting between analytic and experimental results. The symbol rate used in the experiments was $D=28$  $GBaud$. At the transmitter, digital pre-filtering was used to obtain a square-root-raised-cosine (SRRC) spectrum with a roll-off equal to 0.2. The baud rate  $D=28$  $GBaud$ was selected on purpose in order to test the performance of the system without significant opto-electronic bandwidth limitations.

\begin{figure}[h!]
\centering
\includegraphics[width=0.5\textwidth]{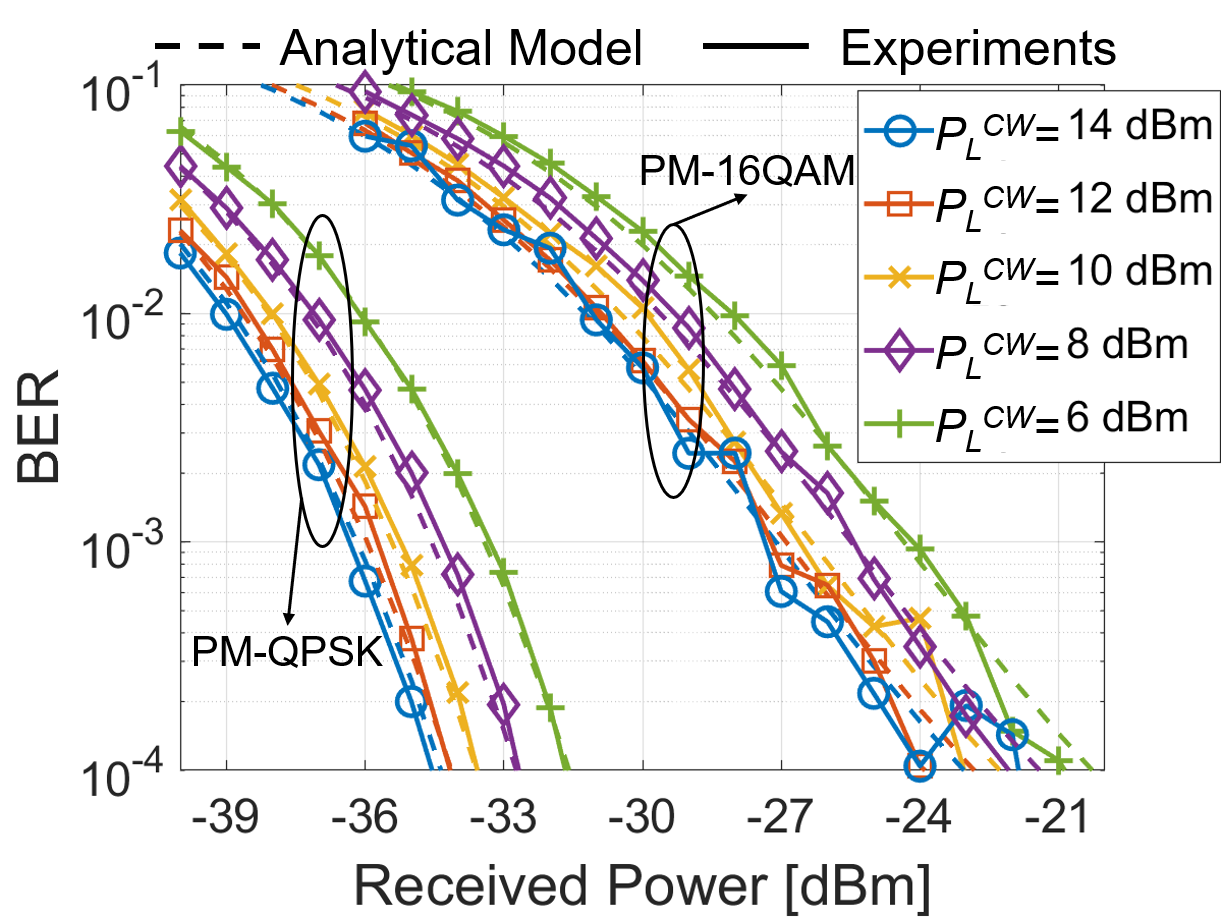}
\caption{BER vs. received signal power for different local oscillator powers $P_L^{CW}$ and two modulation formats (PM-QPSK and PM-16QAM) at 28 $GBaud$. Solid lines: experiments. Dashed lines: analytical model.}
\label{fig:fitting}
\end{figure}

The obtained experimental results are shown in Fig. \ref{fig:fitting} in terms of BER (obtained in off-line processing by error counting on six PRBS repetitions) vs. received average optical power $P_s$ at the input of the coherent receiver, for several LO powers and for the two modulation formats (PM-QPSK and PM-16QAM). On the same figure, we superimpose the curves obtained by the analytical model expressed by Eq.(\ref{eq_SNR}), (\ref{BER_QPSK}), (\ref{BER_16QAM}), in which the values of the four free parameters ($R$, $CMRR$, $i_{TIA}$ and $SNR_Q$) were obtained by a least mean square numerical fitting running simultaneously on all experimental curves shown in Fig. \ref{fig:fitting}. 
The resulting numerically fitted values are shown in Table \ref{table_fitting}, and give a very good agreement with the experimental BER sensitivity curves for all 10 curves reported in Fig. \ref{fig:fitting}. The rest of the paper is based on this excellent matching between experiments and analytical model, allowing us to derive in next Sections general scaling laws for non-amplified coherent receivers.
We also set $RIN=-145$ $dB/Hz$ (as specified in the used laser datasheet), while for the overall receiver noise bandwidth we set $B_{eq}^{RX}=0.6\cdot D$, where  $D$ is the system baud rate. This is a good estimation of the equivalent bandwidth for the full receiver, including the DSP adaptive equalizer. In our experiments at 28 $Gbaud$, we verified that, even though the coherent receiver bandwidth is 22 $GHz$, then the adaptive equalizer (running at two samples per symbol) converges in our experiments to an optimal bandwidth  approximately equal to $B_{eq}^{RX}=0.6\cdot D$.

The parameters optimization  yielded a set of consistent parameters shown in Table \ref{table_fitting}. For instance, the responsivity value $R=0.07$ $A/W$ is perfectly in line with our coherent receiver datasheet, which declares values greater than $R=0.04$ $A/W$ (we remind that for coherent receivers this parameter is not the usual photodiode responsivity, but it is an equivalent responsivity that also takes into account the loss introduced by the optical passive elements placed before the photodiode, i.e. polarizing beam splitters and optical hybrid). Similarly, the fitting on the IRND parameter gives $i_{TIA}=$ 19 $pA/\sqrt{Hz}$, which is again a typical value for high speed TIAs. Parameters in Table \ref{table_fitting} will be used as the starting point for the scalability analysis introduced in the following Section, whose goal is to determine accurate scaling laws for future coherent detection-based short-reach systems on a much larger set of values compared to the one we were able to span in our experiments.

\begin{table}[!h]
%% increase table row spacing, adjust to taste
\renewcommand{\arraystretch}{1.3}
\caption{Model parameters extracted through fitting.}
\label{table_example}
\centering
\begin{tabular}{ccc}
\hline
\textbf{Parameter} & \textbf{Value} & \textbf{Unit}\\
\hline
$R$ & 0.07 & $A/W$\\

$CMRR$ & -20 & $dB$\\

$i_{TIA}$ & 19 & $pA/\sqrt{Hz}$\\

$SNR_Q$ & 18.4 & $dB$\\
\hline
\end{tabular}
\label{table_fitting}
\end{table}

%%%%%%%%%%%%%%%%%%%%%%%%%%%%%%%%%%%%%%%%%%%%%%%%%%%%%%%%%%%%%%%%%%%%%%%%%%%%%%%%%%%%%%%%%%%%%%%%%%%%%%%%%

\subsection{Insertion loss under modulation}  \label{subsec_voltage_opt}

In the previous Section, we were able to obtain a characterization of an unamplified coherent receiver in terms of BER vs. received optical power. The overall available power budget in an unamplified short-reach system is also strictly related to the average optical power available at the output of the optical modulator at the transmitter which, in turn, for PM-QPSK and PM-16QAM depends on the driving peak-to-peak voltage ($V_{pp}$) applied to the modulator, in particular in relation to the  $V_{\pi}$ of the modulator itself. Hence, for the following analysis we define the ``modulation index" parameter as  $m_{index} = V_{pp}/{V_{\pi}}$.

%\begin{align}
%    m_{index} = \dfrac{V_{pp}}{V_{\pi}}%
%    \label{m_index}
%\end{align}

\begin{figure}[h!]
\centering
\includegraphics[width=0.5\textwidth]{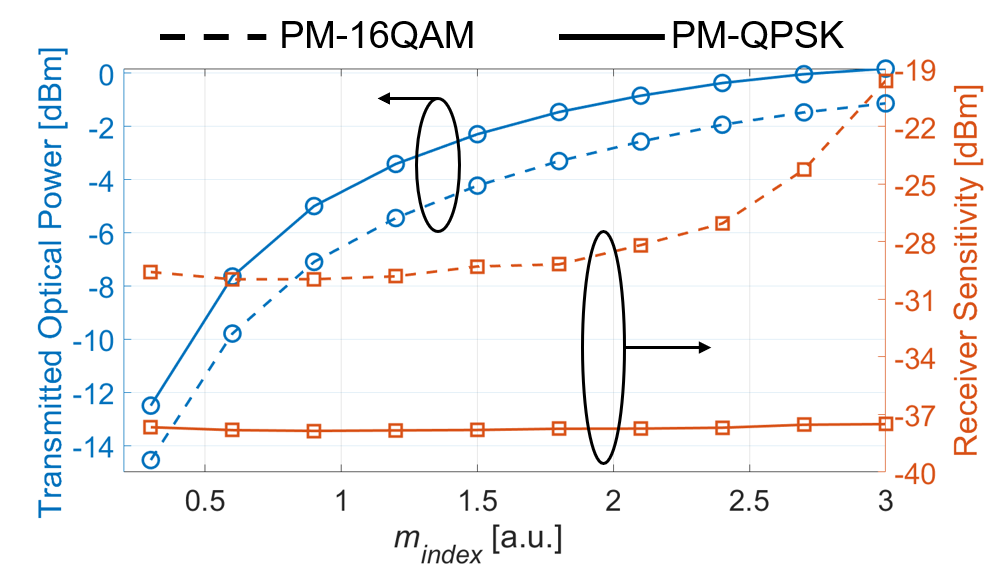}
\caption{Transmitted optical power (left axis labels) and receiver sensitivity $@BER=4\cdot10^{-3}$ (rigth axis labels)  vs.  the ratio between peak-to-peak voltage and $V_{\pi}$. Solid lines: 28 $GBaud$ PM-QPSK modulation. Dashed lines: 28 $GBaud$ PM-16QAM modulation. $P_L^{CW}$= 14 $dBm$.}
\label{fig:Prx_Ptx}
\end{figure}

For a fixed laser CW power=14 $dBm$, Fig. \ref{fig:Prx_Ptx} shows on the left $y$ axis the resulting transmitted optical power at the modulator output, while the right $y$ axis reports the received sensitivity at a BER target level of $4\cdot10^{-3}$, both as a function of the modulation index. As expected, the modulated output power increases with $V_{pp}$, saturating above $2\cdot V_{\pi}$ at the input electrical port of the modulator. 

\begin{figure}[h!]
\centering
\includegraphics[width=0.5\textwidth]{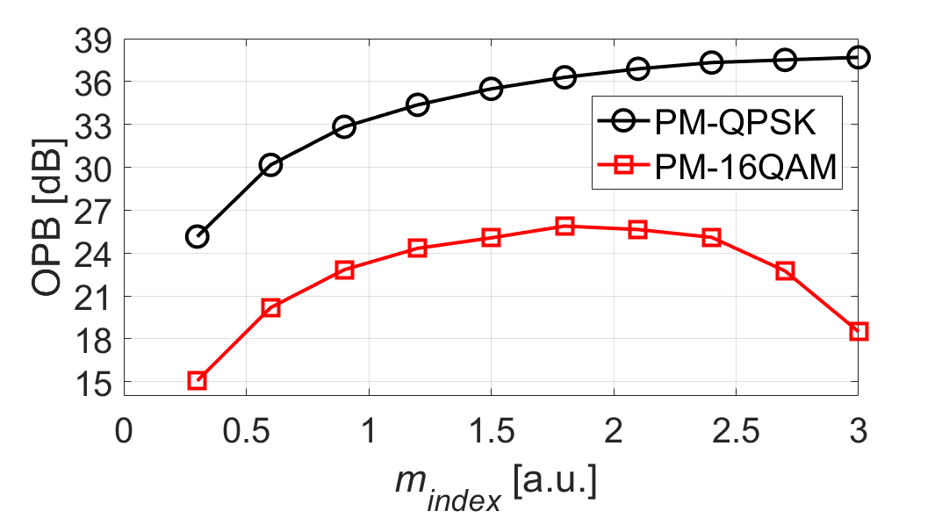}
\caption{Maximum achievable OPB at $BER=4\cdot10^{-3}$ as a function of the ratio between peak-to-peak voltage and $V_{\pi}$. Black curve (circles): 28 $GBaud$ PM-QPSK modulation. Red curve (squares): 28 $GBaud$ PM-16QAM modulation.}
\label{fig:OPB_mindex}
\end{figure}

The received sensitivity remains constant vs. $m_{index}$ in the PM-QPSK case, since there is no significant signal distortion when increasing  $V_{pp}$ inside the analyzed experimental range of Fig. \ref{fig:Prx_Ptx}a, due to the fact that PM-QPSK uses a binary signal on each of the four quadratures of the modulator. On the contrary, for PM-16QAM, the receiver sensitivity starts degrading for high $m_{index}$, since the 4-PAM signal to be applied to each of the four quadratures in the PM-16QAM case becomes distorted when the modulation swing becomes comparable to $2\cdot V_{\pi}$ (we did not use any transmitter non-linearity pre-compensation in our experiments).

The following Fig. \ref{fig:OPB_mindex} re-interprets the results shown in Fig. \ref{fig:Prx_Ptx} in terms of the achievable OPB, defined here as the difference between the modulator output power and the sensitivity of the receiver. For PM-16QAM, an optimum is evident around $m_{index}=2$, while there is not yet an optimum for PM-QPSK in the spanned $V_{pp}$ range up to 3 $V_{\pi}$, even though a saturation is evident, and no further gain would be expected for even higher (and by the way impractical)  $V_{pp}$. For PM-QPSK, performance keeps (slightly) improving above $m_{index}=2$ since for this modulation format the electrical signal on each of the four quadratures is a two-level signal. Thus, in the presence of mild bandwidth limitations in the opto-electronics (electrical driver and optical modulator), as it happens in our experiments, it is known that overdriving the external modulator cleans the binary eye diagrams, since it compresses the signal imperfections in the high and low levels. The graphs in Fig. 4 can be useful for a hardware transceiver designer to find the best trade off in terms of required voltage swing at the input of the modulator between the cost of the electronic driver and the achievable OPB performances.

%\begin{figure}[h!]
%\centering
%\includegraphics[width=0.5\textwidth]{OPB_vs_Vpp_Comparison.jpg}
%\caption{Maximum achievable OPB at $BER=4\cdot10^{-3}$ as a function of the ratio between peak-to-peak volatge and $V_{\pi}$. Black curve (circles): 28 $GBaud$ PM-QPSK modulation. Red curve (squares): 28 $GBaud$ PM-16QAM modulation.}
%\label{fig:ODN_Opt}
%\end{figure}

In order to make some considerations on the best and a sub-optimal case scenarios, the experimental results presented in Section \ref{sec_experimental_validation} have been obtained for only two $m_{index}$ values for the two modulation formats, namely 0.9 and 2.7 for PM-QPSK, and 0.9 and 1.5 for PM-16QAM. 
%%The optimal $m_{index}$ for both modulations has been chosen slightly lower than the actual value to ease some pressure on the modulator.

%%%%%%%%%%%%%%%%%%%%%%%%%%%%%%%%%%%%%%%%%%%%%%%%%%%%%%%%%%%%%%%%%%%%%%%%%%%%%%%%%%%%%%%%%%%%%%%%%%%%%%%%%

\section{Analytical Results} \label{sec:sec_analytical_results}
In this Section, we first analyze the performance of ultra-high-speed coherent short-reach systems at different baud rates and then we study their scalability under some realistic assumptions. We consider all the noise contributions described and experimentally measured in previous Sections, while we do not assume any significant bandwidth limitation when we scale the baud rate up to higher values. The goal of this section is to find ultimate scaling laws in terms of achievable bit rates and OPB limited by intrinsic noises in an unamplified coherent receiver, but not limited by opto-electronic components bandwidth.

\begin{figure}[h!]
\centering
\includegraphics[width=0.5\textwidth]{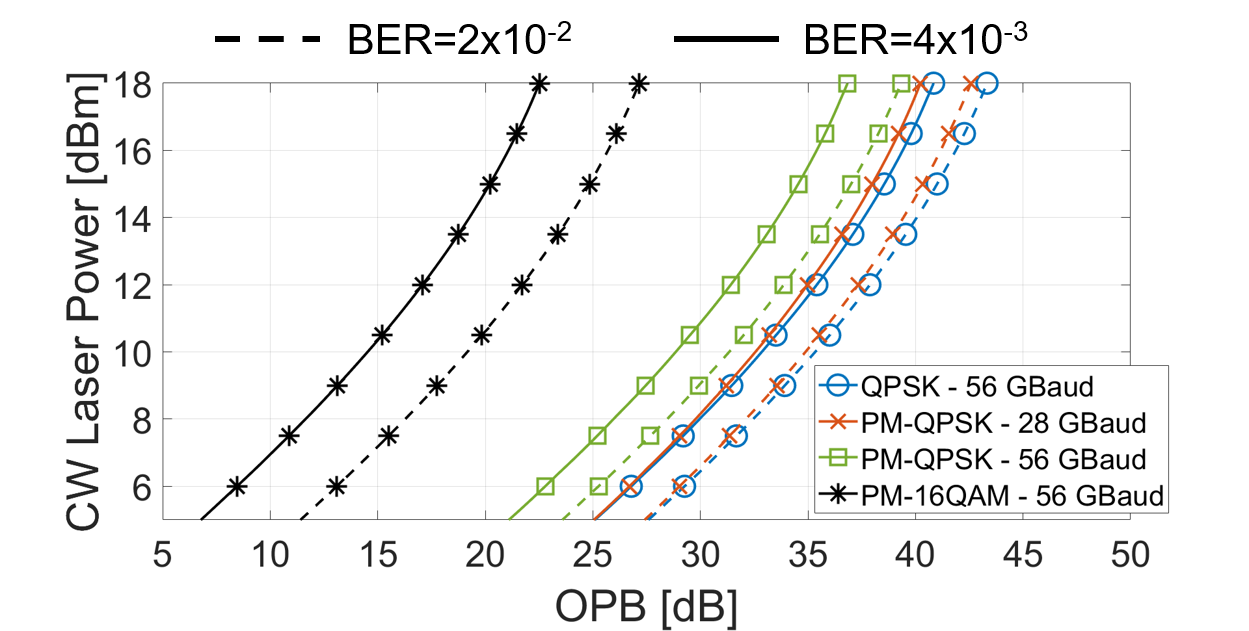}
\caption{Maximum achievable OPB as a function laser power $P_L^{CW}$ for the four cases under consideration. Solid lines: $BER = 4 \cdot 10^{-3}$ (HD-FEC). Dashed lines: $BER = 2 \cdot 10^{-2}$ (SD-FEC).}
\label{fig:Plo_vs_OPB}
\end{figure}

\begin{figure}[h!]
\centering
\includegraphics[width=0.5\textwidth]{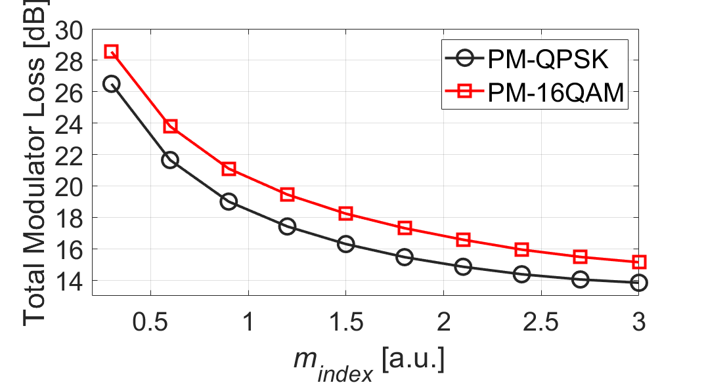}
\caption{Attenuation introduced by the modulator as a function of the ratio between peak-to-peak voltage $V_{pp}$ and $V_{\pi}$. Black curve (circles): PM-QPSK modulation. Red curve (squares): PM-16QAM modulation.}
\label{fig:Mod_Loss}
\end{figure}

Fig. \ref{fig:Plo_vs_OPB} shows the tolerable OPB as a function of the laser CW power $P_L^{CW}$. We assume the same laser technology is used both for the LO and the transmission laser, therefore $P_L^{CW}$ corresponds to the LO power and to the modulator input power as well. Fig. \ref{fig:Plo_vs_OPB} reports the resulting curves at two different BER targets: $BER_{target}=4 \cdot 10^{-3}$ assuming the receiver is equipped with a hard-decision forward error correction (HD-FEC) algorithm, and $BER_{target}=2 \cdot 10^{-2}$ for a more advanced soft-decision FEC (SD-FEC). 
The actual transmitted power is determined by the total modulator loss under modulation shown in Fig. \ref{fig:Mod_Loss}, which in turn was derived from  Fig. \ref{fig:Prx_Ptx}. The overall loss introduced by the modulator is intended as the sum of its passive intrinsic insertion loss and the attenuation associated with the modulation process, which depends on the modulation index. At the considered optimal voltages, the measured modulator loss is 18.2 $dB$ and 14 $dB$ for PM-16QAM and PM-QPSK modulation respectively.

We have investigated four different configurations:
\begin{enumerate}
	\item Dual polarization PM-16QAM at 56 $GBaud$, 448 $Gbps$ gross bit rate, for 400G.
	\item Dual polarization PM-QPSK at 56 $GBaud$, 224 $Gbps$ gross bit rate, for 200G.
	\item Dual polarization PM-QPSK at 28 $GBaud$, 112 $Gbps$ gross bit rate, for 100G.
    \item Single polarization QPSK at 56 $GBaud$, 112 $Gbps$ gross bit rate, for 100G. In this case the modulator loss was set to 10 $dB$.
\end{enumerate}

%\begin{figure}[h!]
%\centering
%\includegraphics[width=0.5\textwidth]{ModulatorLoss_vs_Vpp_Comparison5.jpg}
%\caption{Attenuation introduced by the modulator as a function of the ratio between peak-to-peak voltage $V_{pp}$ and $V_{\pi}$. Black curve (circles): PM-QPSK modulation. Red curve (squares): PM-16QAM modulation.}
%\label{fig:Mod_Loss}
%\end{figure}

According to our model, with the exception of PM-16QAM modulation at 56 $GBaud$ discussed later, in all the other cases coherent systems can provide OPB in excess of, for instance, the 29 $dB$ target required by the ITU-T optical distribution network class N1, for laser powers $P_L^{CW}$ as low as 10 $dBm$ and for both BER targets. Increasing laser power enables even higher OPB classes, such as the 31 $dB$  of Class N2. Assuming an even higher, but still practical laser power of 16 $dBm$, the maximum admissible OPB would be greater than 36 $dB$, which would allow for an extra system margin to account for other system impairments not considered in this work, such as phase noise, burst mode implementation or component imperfections.

Among the four considered configurations, PM-16QAM at 56 $GBaud$ for 400G scenario is the only one that is critical in terms of OPB for PON applications.  In fact, even assuming $P_L^{CW}$=16 $dBm$, the highest achievable OPB is about 27 $dB$ even using SD-FEC. It should be anyway mentioned that ITU-T and IEEE are considering splitting ratio smaller that 64, and thus a lower OPB target, in future architectures for fronthauling PON. Moreover, 23 $dB$ OPB can ensure successful 400G transmission in most of the short-reach system configurations, including inter- and intra-data center, which have much less demanding OPB requirements than PON, but target even higher bit rates.

\begin{figure}[h!]
\centering
\includegraphics[width=0.5\textwidth]{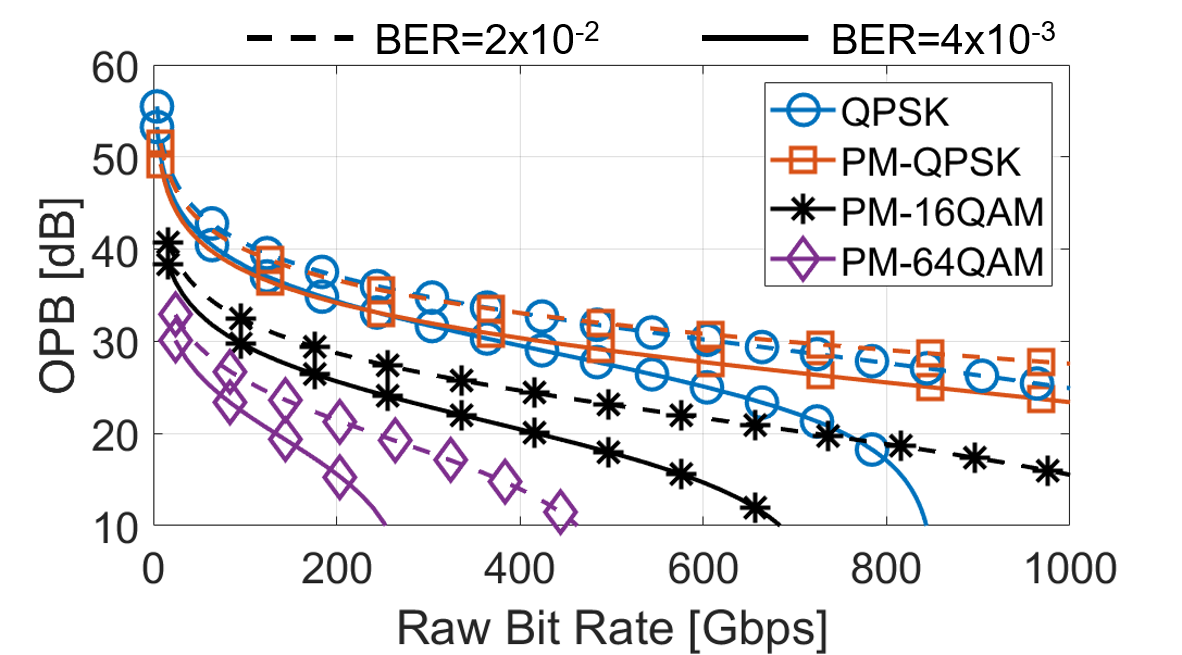}
\caption{OPB as a function of the raw bit rate for three modulation formats. Solid lines: $BER = 4 \cdot 10^{-3}$ (HD-FEC). Dashed lines: $BER = 2 \cdot 10^{-2}$ (SD-FEC). The laser power is set to $P_L^{CW}$= 14 $dBm$.}
\label{fig:ODNLoss_vs_RawBitRate}
\end{figure}

In order to get an even more general view, we show in Fig. \ref{fig:ODNLoss_vs_RawBitRate} the achievable OPB as a function of the raw bit rate at the two BER targets, when laser power $P_L^{CW}$ is set to 14 $dBm$, for four modulation formats including a more advanced PM-64QAM. In terms of general scalability laws of unamplified coherent, this is the most important graph of our paper. It should be remembered that here we did not consider any bandwidth limitation even at extremely high baud rate, so Fig. \ref{fig:ODNLoss_vs_RawBitRate} is to be interpreted as a quite ideal scaling laws curve. In fact, the right end side of the graph would require baud rate (and thus available analog bandwidth) that are for the moment impractical, even though some pioneering experiments are today showing baud rates going towards 100 $GBaud$  \cite{Zhalehpour}. 

Assuming the typical 29 $dB$ OPB target for PON, Fig. \ref{fig:ODNLoss_vs_RawBitRate} shows that QPSK modulation in dual-polarization configuration seems to be the best option, even at extremely high raw bit rates up to 800 $Gbps$ and with simpler, less power-hungry HD-FEC algorithm. On the other hand, PM-16QAM transmission can be taken into consideration only for bit rates up to 190 $Gbps$ and only when very tight system constraints can be tolerated, such as the use of SD-FEC algorithm and no additional system margin. Increasing $P_L^{CW}$ to 16 $dBm$ would allow for bit rates up to approx. 260 $Gbps$ at 29 $dB$ OPB.

In DCI transmission, the required OPB is sensibly lower than in PON, so that all the investigated modulation formats offer ample margin to allow for high-speed communication. For instance, Fig. \ref{fig:ODNLoss_vs_RawBitRate} shows that PM-64QAM can enable high raw bit rates up to 400 $Gbit/s/\lambda$ while still ensuring OPB above 14 $dB$, which could be allocated to take into account the losses due to physical phenomena (e.g. propagation, connectors, nonlinear effects), but also the penalty introduced by novel techniques for power consumption reduction such as DSP-free operations \cite{KAHN} or self-coherent detection \cite{PLANT}.

In both cases, PM-16QAM and PM-64QAM would be interesting since for the same bit rate as PM-QPSK they require significantly lower baud rate and, consequently, lower analog bandwidth in the transceivers.

All the OPB curves at the HD-FEC threshold in Fig. \ref{fig:ODNLoss_vs_RawBitRate} show an inflection point at high bit rates where the system performance quickly degrades for increasing bit rate. This is due to the $SNR_Q$ parameter in Eq. (\ref{eq_SNR}) which produces a floor in the sensitivity curves of BER as a function of the received signal optical power. In our model, we extracted $SNR_Q$ through fitting with the experimental curves at 28 $GBaud$ and then scaled it inversely to the baud rate to match the conditions of our experimental measurements. In fact, our DAC and ADC run at fixed sampling rates (92 and 200 $Gsample/s$ respectively) but the DSP then applies a downsampling to two samples per symbol, effectively low-pass filtering the received signal and thus progressively reducing the variance of quantization noise for increasing baud rates. It is worth mentioning that this would not be the case in a realistic installed system, where DACs and ADCs are usually chosen in accordance to the baud rate the system has to work at.

\begin{figure}[h!]
\centering
\includegraphics[width=0.5\textwidth]{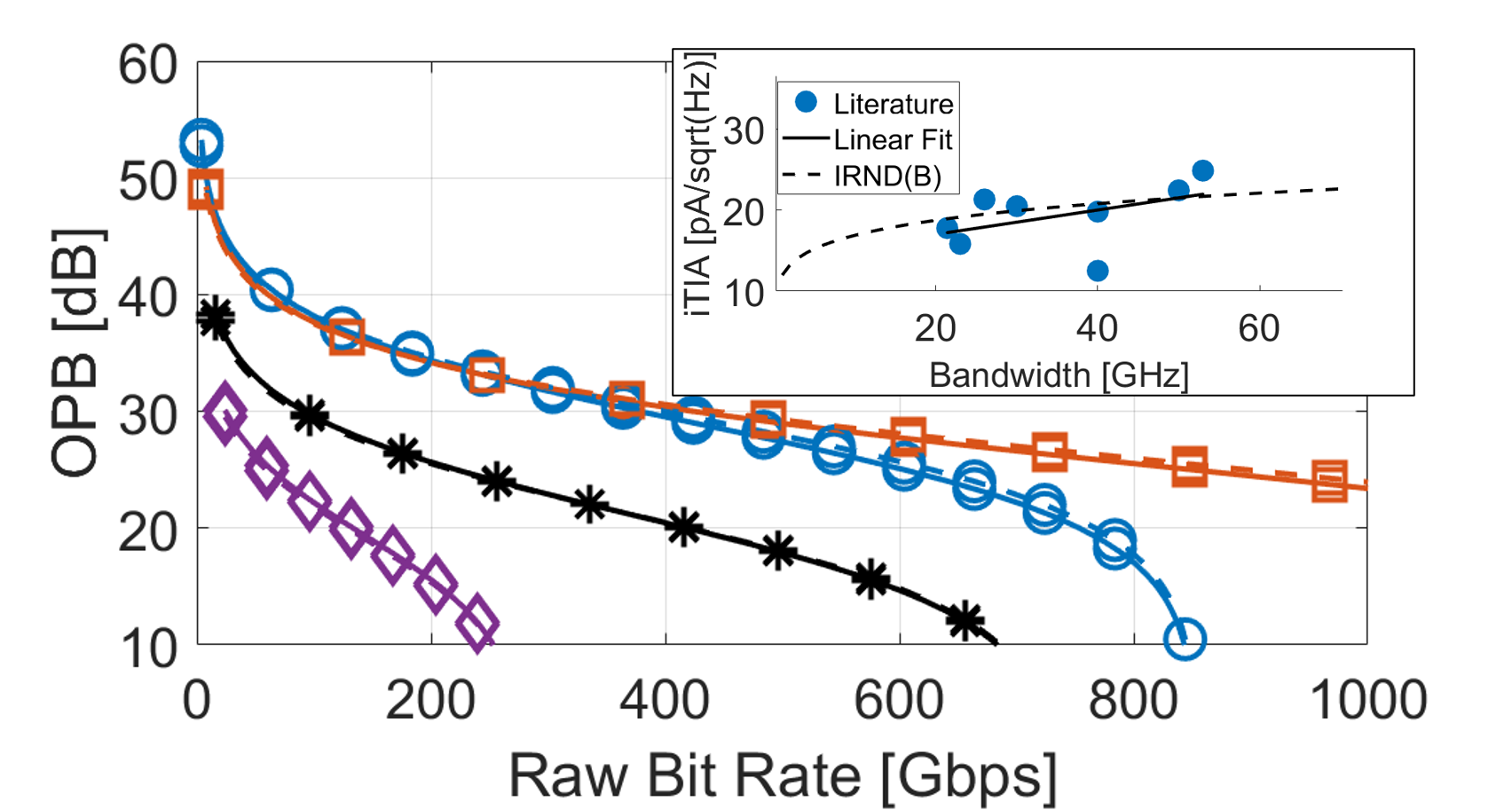}
\caption{OPB as a function of the raw bit rate for three modulation formats. Solid lines: with $i_{TIA}(B)=i_{TIA}(B_0)*(B/B_0)^x$. Dashed lines: with $i_{TIA}(B)=$ 19 $pA/\sqrt{Hz}$. The laser power is set to $P_L^{CW}$= 16 $dBm$. $BER = 4 \cdot 10^{-3}$ (HD-FEC). Legend in Fig. \ref{fig:ODNLoss_vs_RawBitRate} applies also here.}
\label{fig:ODNLoss_vs_RawBitRate_tia}
\end{figure}

Another important parameter in a coherent receiver is the IRND $i_{TIA}$ that represents the TIA-induced thermal noise contribution. The results in Fig.\ref{fig:ODNLoss_vs_RawBitRate} were obtained for a fixed $i_{TIA}=$ 19 $pA/\sqrt{Hz}$ value extracted through the fitting procedure. However, literature results suggest that IRND tends to slightly increase for higher bandwidth TIA. Since we want to study in this paper the scalability for a wide range of bandwidths, we further investigate on this point. A detailed literature review of transimpedance amplifiers performance including noise current spectral density and bandwidth can be found in \cite{TIA} for different technologies and architectures. Starting from the devices listed in Table I in \cite{TIA}, and after removing two "outlier" values, we have extrapolated through linear polynomial fitting a relation between the bandwidth of a TIA and its noise current density (the $i_{TIA}$ parameter) based on the state of the art. The inset in Fig. \ref{fig:ODNLoss_vs_RawBitRate_tia} shows the parameter for the devices reported in \cite{TIA} (blue dots), the result of a linear polynomial fitting on the set of published performance (solid line) and the curve (dashed line) describing the relation $i_{TIA}(B)=i_{TIA}(B_0)*(B/B_0)^x$ where $B$ is the TIA bandwidth, $B_0$ is the bandwidth of our coherent receiver (22 $GHz$), $i_{TIA}(B_0)$ is the TIA noise current density extracted through fitting in section \ref{sec:Exp} (19 $pA/\sqrt{Hz}$) and $x$ is the polynomial coefficient returned by the linear fitting procedure.

A comparison of OPB vs raw bit rate curves obtained with a variable $i_{TIA}(B)$ parameter and with the constant $i_{TIA}(B)=$ 19 $pA/\sqrt{Hz}$ is shown in Fig. \ref{fig:ODNLoss_vs_RawBitRate_tia}. The two sets of curves are almost perfectly superimposed highlighting that, at least for high values of the LO power (16 $dBm$ in this case), the effect of the thermal noise variation with the receiver bandwidth is practically negligible, since the system performance for increasing bit rate is limited by other noise sources. Anyway, for lower LO power our proposed scaling law for $i_{TIA}(B)$ can have a practical relevance.

%%%%%%%%%%%%%%%%%%%%%%%%%%%%%%%%%%%%%%%%%%%%%%%%%%%%%%%%%%%%%%%%%%%%%%%%%%%%%%%%%%%%%%%%%%%%%%%%%%%%%%%%%

\section{Experimental Validation on an installed fiber metropolitan testbed} \label{sec_experimental_validation}

An experimental confirmation of the results depicted in Fig. \ref{fig:Plo_vs_OPB} and Fig. \ref{fig:ODNLoss_vs_RawBitRate} can be obtained by comparison with Fig. \ref{fig:OPB_mindex}. For PM-QPSK at $m_{index}=2.7$ (the considered optimal modulation index) the OPB at $BER = 4 \cdot 10^{-3}$ is 37.51 $dB$  in Fig. \ref{fig:OPB_mindex}, whereas it is 37.13 $dB$ both from Fig. \ref{fig:Plo_vs_OPB} (solid red curve at 14 $dBm$ LO power) and From Fig. \ref{fig:ODNLoss_vs_RawBitRate} (solid red curve at 112 $Gbps$). For PM-16QAM at $m_{index}=1.5$ the OPB at $BER = 4 \cdot 10^{-3}$ is 25 $dB$ in Fig. \ref{fig:OPB_mindex} and 24.81 $dB$ in Fig. \ref{fig:ODNLoss_vs_RawBitRate} at 224 $Gbps$. Slight discrepancies can be due to a non perfect parameter fitting, but also to a slightly different LO power in the experiments as opposed to the ideal power considered in the model.

\begin{figure}[h!]
\centering
\includegraphics[width=0.5\textwidth]{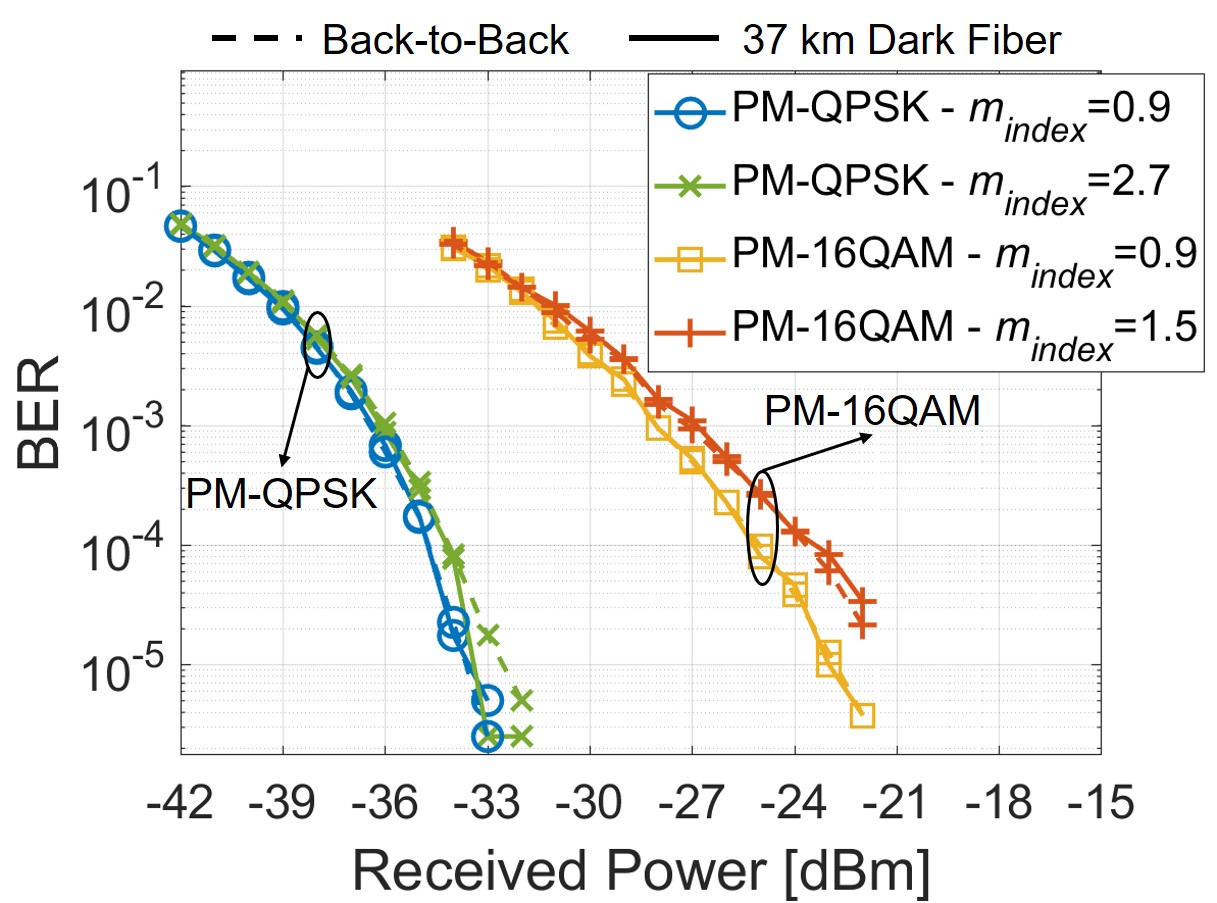}
\caption{Comparison of sensitivity curves in back-to-back (dashed) and with transmission over 37 km of installed dark fiber (solid), for the two modulation formats at 28 $GBaud$ and for the best and a sub-optimal modulation index.}
\label{fig:Fastweb}
\end{figure}

Results depicted in Fig. \ref{fig:ODNLoss_vs_RawBitRate} have also been experimentally verified in a realistic short-reach scenario for both PM-QPSK and PM-16QAM modulation in back-to-back transmission and over a 37 $km$ installed metropolitan dark fiber optical link. We could only perform measurements at 28 $GBaud$ due to available hardware. The results are shown in Fig. \ref{fig:Fastweb}. For both modulations no significant impairment can be observed when a 37 $km$ fiber link is considered, with respect to the back-to-back condition. The sensitivity (solid) curves are perfectly superimposed on the back-to-back (dashed) curves, showing negligible detrimental effects due to propagation in the fiber. Moreover, the coherent receiver sensitivity is affected by the modulation peak-to-peak voltage only to a small extent. Nevertheless, as already stated in Section \ref{subsec_voltage_opt}, the decrease in sensitivity is counterbalanced by the reduced insertion loss of the modulator, which results in an overall better performance in terms of achievable OPB.

%%%%%%%%%%%%%%%%%%%%%%%%%%%%%%%%%%%%%%%%%%%%%%%%%%%%%%%%%%%%%%%%%%%%%%%%%%%%%%%%%%%%%%%%%%%%%%%%%%%%%%%%%

\section{Examples of Practical Implementation of Our Model} \label{implementation}
\begin{figure}[h!]
\centering
\includegraphics[width=0.5\textwidth]{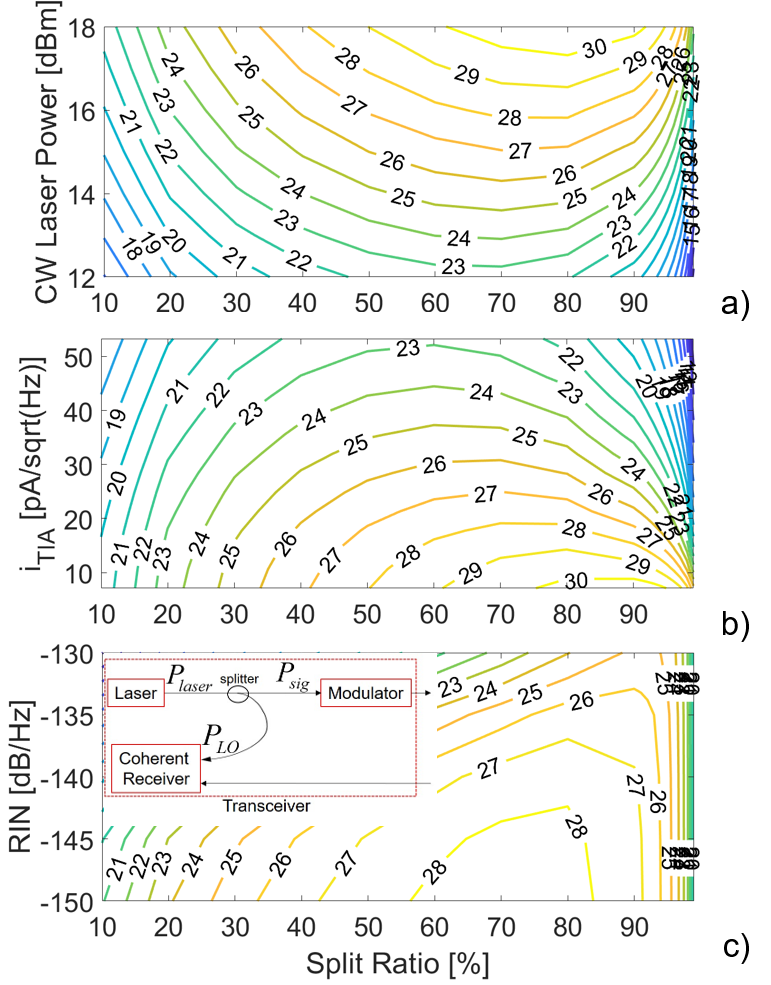}
\caption{a) OPB as a function of split ratio and CW laser power. b) OPB as a function of split ratio and TIA noise current density for 16 $dBm$ laser power. c) OPB as a function of split ratio and laser RIN for 16 $dBm$ laser power. 28 $GBaud$ PM-16QAM modulation at $BER = 2 \cdot 10^{-2}$ (SD-FEC).}
\label{fig:contour}
\end{figure}

In this Section, we show as an example the application of our model to predict and optimize the performance of specific transceiver architectures for inter-data center coherent transceivers. For instance, some recently developed integrated transceivers are equipped with a single laser followed by a splitter that feeds the signal modulator and the local oscillator \cite{Transceiver} as shown in the inset of Fig. \ref{fig:contour}c. In the following, we use our model to predict system performance of this architecture and to optimize the splitting ratio after the laser. Moreover, to extend the scope of the paper to a wider range of possible component parameters, we show in the folllowing three graphs for the specific case of 28 $GBaud$ PM-16QAM modulation at the SD-FEC threshold, the achievable OPB when changing three of the most relevant system parameters: laser power, TIA noise current and RIN. We define the split ratio as $\rho = P_{sig}/{P_{laser}}$, where $P_{sig}$ is the signal power at the external modulator input and $P_{laser}$ is the total power emitted by the laser before the splitting stage (Fig. \ref{fig:contour}c).

The contour plot of OPB as a function of the split ratio and of the CW laser power is depicted in Fig. \ref{fig:contour}a. The optimal split ratio strongly depends on $P_{laser}$ starting from 65\% for low $P_{laser}$ and moving towards 80\% for very high $P_{laser}$, suggesting that the impact of increased RIN contribution has to be mitigated by re-balancing the signal and LO power at the receiver (see Eq. (\ref{eq_SNR})). For a 16 $dBm$ CW laser power the achievable OPB is about 28 $dB$ with a split ratio in the range 70-80\%.

The effect of different IRND values on the optimal split ratio is shown in Fig. \ref{fig:contour}b for a 16 $dBm$ CW laser power. For low $i_{TIA}$ values, the optimal ratio is again in the order of 80\%, but reduces down to about 60\% when the thermal noise becomes dominant. Finally, Fig. \ref{fig:contour}c shows how RIN affects the optimal split ratio and the overall OPB. As expected, when the RIN coefficient is low, a stronger LO is advantageous, whereas for high RIN values the optimal split ratio moves towards a higher percentage of signal power. Similar results were obtained (but not shown for space limitations), and therefore similar conclusions can be drawn for the other three modulation formats under analysis, regardless of the selected FEC threshold.

%%%%%%%%%%%%%%%%%%%%%%%%%%%%%%%%%%%%%%%%%%%%%%%%%%%%%%%%%%%%%%%%%%%%%%%%%%%%%%%%%%%%%%%%%%%%%%%%%%%%%%%%%

\section{Discussion and Conclusion}

In the mid-to-long-term timeframe assumed in our analysis, several techno-economic factors will affect the decision on whether or not coherent technologies will replace DD for short-reach. It seems anyway unlikely that IM-DD systems will be able to cope with the growing traffic demand well beyond 100G$/\lambda$ without the need to sensibly increase their complexity and cost \cite{Pablo}. In our opinion, a turning point can be 400G per wavelength for short reach data center solutions, and 100G+ for access since at this data rate chromatic dispersion and power budget become exceedingly critical for DD. Current IM-DD higher speed standards, for instance, envisage multiple lanes (wavelength or parallel fibers) and/or optical amplification to ensure the same aggregate bit rate provided by a single-channel coherent system, resulting in additional hardware components coupled with some form of DSP both at the transmitter and receiver side, which may reduce the cost-effectiveness of DD solutions.

Moreover, extensive research is today being performed on hardware and software solutions targeting power/cost reduction of coherent technology in the mid-to-long-term. Studies on self-coherent systems \cite{PLANT2,Gui}, DSP-free configurations \cite{KAHN}, low-power DSP \cite{Freund}, optical signal processing \cite{PLANT2,KAHN}, narrow-linewidth semiconductor low-cost lasers \cite{Huang} and integration \cite{Brodnik} are achieving ground-breaking results that will likely pave the way to the introduction of coherent in short reach communications. Some specific techno-economic studies tailored for 400G have shown that a single-wavelength coherent solution has very similar power consumption and ASIC footprint of a 4 lanes 100G DD solution \cite{Cheng}.

Given the ongoing ASIC power consumption and complexity improvements \cite{Cheng,Xie,Zhou}, one of the remaining issues related to the transition from DD to coherent is related to the quality of the lasers in terms of low linewidth for phase noise requirements. Currently, coherent transmission normally uses costly external cavity lasers ($<$100 kHz linewidth). However, chances are that as technology improves, low-cost lasers with these features will become available. Moreover, at least for QPSK modulation, experiments have already shown the possibility of using lower cost DFB lasers \cite{Rizzelli}.

The tolerance to laser linewidth is, thus, one of the major issues to be solved to enable the use of coherent for short reach communications. Nevertheless, recent works show:

\begin{itemize}
   \item For QPSK modulation the tolerance is very high. Total linewidths up to tens of MHz can be allowed by using appropriate phase estimation methods such as maximum a posteriori or Wiener filter-based \cite{Taylor}.
   \item A 1 $dB$  SNR penalty is achieved at the soft-FEC threshold in 32 $GBaud$  coherent systems when the total linewidth (transmission laser + local oscillator) is 9.6 $MHz$ or 1.8 $MHz$ for 16-QAM and 64-QAM modulation respectively, by using a non-linear least squares method \cite{Argyris}.
   \item The total linewidth tolerance at 1 $dB$ SNR penalty and BER=$10^{-3}$ for a 32 $GBaud$ PM-16QAM system is about 4 $MHz$ when a blind phase search based algorithm is used \cite{ke}.
\end{itemize}

Thus, the laser linewidth seems to be of great concern only for higher order modulation formats such as 64-QAM, provided that a proper carrier phase estimation algorithm is used. Moreover, the impact of phase noise reduces with the data rate \cite{Taylor,Argyris,ke}, which helps envisaging the use of low-cost DFB lasers for high-speed short-reach coherent communications thanks to the high available power budget shown in our paper.

A roadmap from traditional DD transceivers to coherent transceivers in short-reach will require a complete revolution in the techno-economics of such systems, mostly in terms of opto-electronic components and on the DSP in the transceivers ASIC chipsets. On the contrary, regarding the protocol stacks, and in particular regarding interoperability issues, the changes for the Physical Media Dependent (PMD) and the Physical Medium Attachment (PMA) sublayers would be relatively limited, even though to be considered in detail. For instance, the initial bootstrap of a coherent receiver will need a relatively long training sequence for the initial settings of the local oscillator central frequency and then for setting the RX adaptive equalizer initial taps condition.

In conclusion, we have given general scaling laws for unamplified ultra-high speed coherent short-reach systems, showing results that may allow vendors and telecom operators in the PON and DCI areas to drive their decision on which bit rate will require a complete paradigm change from DD to coherent in short-reach.

To do so, we have extended a previously introduced \cite{zhang} analytical model for predicting the performance of unamplified full coherent receiver affected by shot, thermal and RIN noise sources by including the influence of additional power-independent implementation factors, through the term $SNR_Q$ in Eq. (1). We have validated our model and optimized its parameters to obtain the best fitting with experimental results for two different modulation formats, PM-16QAM and PM-QPSK. The model-predicted coherent receiver sensitivity has proven to be in excellent agreement with the experimental results. We have also performed an experimental optimization of the modulation peak-to-peak voltage finding the best operating conditions in the linear regime of the electro-optical modulator.

The model has then been used to generate scaling laws for amplifier-less coherent detection-based high-speed short-reach networks showing that 100G and 200G per wavelength are at hand from a power budget perspective, when single- or dual-polarization QPSK modulation is used, providing also a conspicuous additional system margin of more than 8 $dB$ over the typical 29 $dB$ OPB required by modern PON standards.

Our findings also highlight that PM-16QAM modulation could be used to reach 190 $Gbps$ speed at the 29 $dB$ OPB threshold, but only in combination with advanced SD-FEC algorithms and high laser power level. Nonetheless, PM-16QAM seems to be largely suitable for short-reach systems that require less than 23 $dB$ OPB at 400G. In DCI communication, where the OPB requirement can be further relaxed, the use of a more advanced PM-64QAM format can be envisaged up to 800 $Gbit/s/\lambda$  thanks to the $>$14 $dB$ OPB ensured in the considered bit rate range.

We have validated these results with a transmission experiment using PM-QPSK and PM-16QAM modulation at 28 $GBaud$ in back-to-back and on a 37 $km$ fiber link. OPB values notably close to the ones predicted by the model have been obtained and no impact of fiber propagation impairments has been observed.

Lastly, as an example of application we have shown the effect of various system parameters on the optimal split ratio required for practical implementation with a single laser for transmission and reception in the transceiver. 

%%%%%%%%%%%%%%%%%%%%%%%%%%%%%%%%%%%%%%%%%%%%%%%%%%%%%%%%%%%%%%%%%%%%%%%%%%%%%%%%%%%%%%%%%%%%%%%%%%%%%%%%%

% use section* for acknowledgment
\section*{Acknowledgment}

This work was carried out under the PhotoNext initiative at Politecnico di Torino (http://www.photonext.polito.it/). The authors would also like to thank Cisco Photonics for the fruitful discussion exchanged on these topics.

% Can use something like this to put references on a page
% by themselves when using endfloat and the captionsoff option.
\ifCLASSOPTIONcaptionsoff
  \newpage
\fi


\begin{thebibliography}{100}
%\newcommand{\enquote}[1]{``#1''}

\bibitem{Infinera}
Coherent WDM Technologies - White Paper, Infinera Corporation, Sunnyvale, CA, USA, 2016 [Online]. Available: https://www.infinera.com/wp-content/uploads/Infinera\_Coherent\_Tech.pdf.

\bibitem{Winzer}
P.~Winzer, D.~Neilson, and A.~Chraplyvy, ``Fiber-optic transmission and networking: the previous 20 and the next 20 years [Invited],'' Optics Express \textbf{26}, 24190--24239 (2018).

\bibitem{400G-ZR}
The Optical Internetworking Forum, ``Flex Coherent DWDM Transmission Framework Document," August 3, 2017 [Online]. Available: https://www.oiforum.com/wp-content/uploads/2019/01/OIF-FD-FLEXCOH-DWDM-01.0-1.pdf.

\bibitem{Kahn1}
J.~Kahn, and K.~Ho, ``Spectral efficiency limits and modulation/detection techniques for DWDM systems,'' IEEE Journal of Selected Topics in Quantum Electronics \textbf{10}, 259--272 (2004).

\bibitem{mecozzi}
A.~Mecozzi, and M.~Shtaif, ``Information Capacity of Direct Detection Optical Transmission Systems,'' Journal of Lightwave Technology \textbf{36}, 689--694 (2018).

\bibitem{straullu}
S.~Straullu, J.~Chang, R.~Cigliutti, V.~Ferrero, A.~Nespola, A.~Vinci, S.~Abrate, and R.~Gaudino, ``Single-Wavelength Downstream FDMA-PON at 32 Gbps and 34 dB ODN Loss,'' IEEE Photonics Technology Letters \textbf{27}, 774--777 (2015).

\bibitem{IEEE}
IEEE P802.3ca 50G-EPON Task Force, [Online]. Available: http://www.ieee802.org/3/ca/.

\bibitem{ITU}
ITU-T G.Sup64 PON transmission technologies above 10 Gbit/s per wavelength, [Online]. Available: https://www.itu.int/rec/T-RECG. Sup64-201802-I/en.

\bibitem{NGPON2}
J.~Wey, D.~Nesset, M.~Valvo, C.~Grobe, H.~Roberts, Y.~Luo, and J.~Smith, ``Physical layer aspects of NG-PON2 standards—Part 1: Optical link design [Invited],'' IEEE/OSA Journal of Optical Communications and Networking \textbf{8}, 33--42 (2016).

\bibitem{Pilori}
D.~Pilori, L.~Bertignono, A.~Nespola, F.~Forghieri, and G.~Bosco, ``Comparison of Probabilistically Shaped 64QAM With Lower Cardinality Uniform Constellations in Long-Haul Optical Systems,'' Journal of Lightwave Technology \textbf{36}, 501--509 (2018).

\bibitem{Erkilinc}
M.~Erkilin\c{c}, D.~Lavery, K.~Shi, B.~Thomsen, R.~Killey, S.~Savory, and P.~Bayvel, ``Comparison of Low Complexity Coherent Receivers for UDWDM-PONs ($\lambda$-to-the-User),'' Journal of Lightwave Technology \textbf{36}, 3453--3464 (2018).

\bibitem{Matsuda}
K.~Matsuda, R.~Matsumoto, H.~Miura, K.~Onohara, and N.~Suzuki, ``Hardware-Efficient Signal Processing Technologies for Coherent PON Systems,'' European Conference on Optical Communication (2018).

\bibitem{Savory}
S.~J.~Savory, ``Digital Coherent Optical Receivers: Algorithms and Subsystems,'' IEEE Journal of Selected Topics in Quantum Electronics \textbf{16}, 1164--1179 (2010).

\bibitem{XR_optics} A. Rashidinejad, A. Nguyen, M. Olson, S. Hand and D. Welch, ``Real-Time Demonstration of 2.4Tbps (200Gbps/$\lambda$) Bidirectional Coherent DWDM-PON Enabled by Coherent Nyquist Subcarriers,'' 2020 Optical Fiber Communications Conference and Exhibition (OFC), San Diego, CA, USA, 2020, pp. 1-3.

\bibitem{FEC}
E.~de Betou, E.~Mobilon, B.~Angeli, P.~\"Ohlen, A.~Lindstr\"om, S.~Dahlfort, and E.~Trojer, ``Upstream FEC performance in combination with burst mode receivers for next generation 10 Gbit/s PON,'' 36th European Conference and Exhibition on Optical Communication (2010).

\bibitem{Drift}
F.~Saliou, B.~Le Guyader, L.~Guillo, G.~Simon, P.~Chanclou, G.~Bo, G.~Jianhe, and W.~Xuming, ``Upstream wavelength drift during burst time for G-PON, XG-PON1 and TWDM-PON co-existing on the same ODN,'' 2014 The European Conference on Optical Communication (2014).

\bibitem{burst_mode_coherent} R.~Koma, M.~Fujiwara, J.~Kani, K.~Suzuki, and A.~Otaka, ``Burst-mode digital signal processing that pre-calculates FIR filter coefficients for digital coherent PON upstream,'' IEEE/OSA Journal of Optical Communications and Networking \textbf{10}, 461--470 (2018).

\bibitem{kikuchi2}
K.~Kikuchi, ``Fundamentals of Coherent Optical Fiber Communications,'' Journal of Lightwave Technology \textbf{34}, 157--179 (2016).

\bibitem{zhang}
B.~Zhang, C.~Malouin, and T.~Schmidt, ``Design of coherent receiver optical front end for unamplified applications,'' Optics Express \textbf{20}, 3225--3234 (2012).

\bibitem{kikuchi}
K.~Kikuchi, and S.~Tsukamoto, ``Evaluation of Sensitivity of the Digital Coherent Receiver,'' Journal of Lightwave Technology \textbf{26}, 1817--1822 (2008).

\bibitem{lavery}
D.~Lavery, S.~Liu, Y.~Jeong, J.~Nilsson, P.~Petropoulos, P.~Bayvel, and S.~Savory, ``Realizing High Sensitivity at 40 Gbit/s and 100 Gbit/s,'' Optical Fiber Communication Conference, paper OW3H.5 (2012).

\bibitem{Zhalehpour}
S. Zhalehpour et al., ``All-Silicon IQ Modulator for 100 GBaud 32QAM Transmissions,'' Optical Fiber Communications Conference and Exhibition (2019).

\bibitem{KAHN} J. K. Perin, A. Shastri and J. M. Kahn, ``Design of Low-Power DSP-Free Coherent Receivers for Data Center Links,'' IEEE J. Lightwave Technol. \textbf{35}, 4650--4662 (2017).

\bibitem{PLANT} M. Morsy-Osman, M. Sowailem, E. El-Fiky, T. Goodwill, T. Hoang, S. Lessard and D. V. Plant, ``DSP-free coherent-lite transceiver for next generation single wavelength optical intra-datacenter interconnects,''  Optics Express \textbf{26}, 8890--8903 (2018).

\bibitem{TIA} S. G. Kim, C. Hong, Y. S. Eo, J. Kim and S. M. Park, ``A 40-GHz Mirrored-Cascode Differential Transimpedance Amplifier in 65-nm CMOS," in IEEE Journal of Solid-State Circuits,''  IEEE Journal of Solid-State Circuits\textbf{54}, 1468--1474 (2019).

\bibitem{Transceiver} Optical Internetworking Forum - Implementation Agreement for Integrated Coherent Transmit-Receive Optical Sub Assembly (IC-TROSA), 2019 [Online]. Available: https://www.oiforum.com/wp-content/uploads/OIF-IC-TROSA-01.0.pdf.

\bibitem{Pablo} P. Torres-Ferrera, G. Rizzelli Martella, V. Ferrero and R. Gaudino, "100+ Gbps/ 50km C-band Downstream PON using CD Digital Pre-Compensation and Direct-Detection ONU Receiver," Journal of Lightwave Technology.

\bibitem{PLANT2} Mohamed Morsy-Osman, Mohammed Sowailem, Eslam El-Fiky, Tristan Goodwill, Thang Hoang, Stephane Lessard, and David V. Plant, ``DSP-free ‘coherent-lite’ transceiver for next generation single wavelength optical intra-datacenter interconnects,'' Opt. Express \textbf{26}, 8890--8903 (2018).

\bibitem{Gui} T. Gui, X. Wang, M. Tang, Y. Yu, Y. Lu, and L. Li, ``Real-Time Demonstration of 600 Gb/s DP-64QAM SelfHomodyne Coherent Bi-Direction Transmission with Un-Cooled DFB Laser,'' Optical Fiber Communication Conference Postdeadline Papers (2020).

\bibitem{Freund} M.~Erkilin\c{c},, R. Emmerich, K. Habel, V. Jungnickel, C. Schmidt-Langhorst, C. Schubert, and R. Freund, ``PON transceiver technologies for $\geq$50 Gbits/s per lambda: Alamouti coding and heterodyne detection [Invited],'' Journal of Optical Communications and Networking \textbf{12}, A162--A17, 2020.

\bibitem{Huang} Huang, S., Zhu, T., Yin, G. et al. ``Dual-cavity feedback assisted DFB narrow linewidth laser,'' Scientific Reports \textbf{7}, 1185 (2017).

\bibitem{Brodnik} G. M. Brodnik, M. W. Harrington, D. Bose, A. M. Netherton, W. Zhang, L. Stern, P. A. Morton, J. E. Bowers, S. B. Papp, and D. J. Blumenthal, ``Chip-Scale, Optical-Frequency-Stabilized PLL for DSP-Free, Low-Power Coherent QAM in the DCI,'' Optical Fiber Communication Conference, paper M3A.6 (2020).

\bibitem{Cheng} J. Cheng, C. Xie, Y. Chen, X. Chen, M. Tang and S. Fu, ``Comparison of Coherent and IMDD Transceivers for Intra Datacenter Optical Interconnects,'' 2019 Optical Fiber Communications Conference and Exhibition (2019).

\bibitem{Xie} C. Xie and J. Cheng, ``Coherent Optics for Data Center Networks,'' 2020 IEEE Photonics Society Summer Topicals Meeting Series (SUM) (2020).

\bibitem{Zhou} X. Zhou, R. Urata and H. Liu, ``Beyond 1 Tb/s Intra-Data Center Interconnect Technology: IM-DD OR Coherent?,'' Journal of Lightwave Technology \textbf{38}, 475--484 (2020).

\bibitem{Rizzelli} R. Gaudino, V. Curri, G. Bosco, G. Rizzelli, A. Nespola, D. Zeolla, S. Straullu, S. Capriata, and P. Solina, "On the use of DFB Lasers for Coherent PON," Optical Fiber Communication Conference, paper OTh4G.1 (2012).

\bibitem{Taylor} M. G. Taylor, ``Phase Estimation Methods for Optical Coherent Detection Using Digital Signal Processing,'' Journal of Lightwave Technology \textbf{27}, 901--914 (2009).

\bibitem{Argyris} N. Argyris, S. Dris, C. Spatharakis and H. Avramopoulos, ``High performance carrier phase recovery for coherent optical QAM,'' 2015 Optical Fiber Communications Conference and Exhibition (2015).

\bibitem{ke} J. H. Ke, K. P. Zhong, Y. Gao, J. C. Cartledge, A. S. Karar and M. A. Rezania, ``Linewidth-Tolerant and Low-Complexity Two-Stage Carrier Phase Estimation for Dual-Polarization 16-QAM Coherent Optical Fiber Communications,'' Journal of Lightwave Technology \textbf{30}, 3987--3992 (2012).


\end{thebibliography}
\end{document}